# One-Dimensional Collision Carts Computer Model and its Design Ideas for Productive Experiential Learning

Loo Kang WEE

Ministry of Education, Education Technology Division, Singapore
weelookang@gmail.com , wee_loo_kang@moe.gov.sg

Abstract: We develop an Easy Java Simulation (EJS) model for students to experience the physics of idealized one-dimensional collision carts. The physics model is described and simulated by both continuous dynamics and discrete transition during collision. In the field of designing computer simulations, we discuss briefly three pedagogical considerations such as 1) consistent simulation world view with pen paper representation, 2) data table, scientific graphs and symbolic mathematical representations for ease of data collection and multiple representational visualizations and 3) game for simple concept testing that can further support learning. We also suggest using physical world setup to be augmented complimentarily with simulation while highlighting three advantages of real collision carts equipment like tacit 3D experience, random errors in measurement and conceptual significance of conservation of momentum applied to just before and after collision. General feedback from the students has been relatively positive, and we hope teachers will find the simulation useful in their own classes.

2015 Resources added: http://iwant2study.org/ospsg/index.php/interactive-resources/physics/02-newtonian-mechanics/02-dynamics/46-one-dimension-collision-js-model

http://iwant2study.org/ospsg/index.php/interactive-resources/physics/02-newtonian-mechanics/02-dynamics/195-elastic-collision

PACS: 45.10.-b   45.20.df   45.20.dh   01.40.-d   01.50.H-

## I. INTRODUCTION

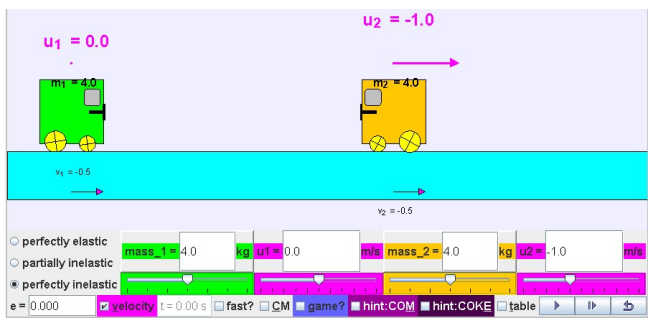

Figure 1.   EJS applet view of the virtual laboratory simulation learning environment showing a world view and a bottom control panel for student-directed inquiry activities .

Conservation of linear momentum coupled with situations when kinetic energy is conserved in perfectly elastic while not during inelastic collisions, are difficult concepts (Singh & Rosengrant, 2003) for many students. This is probably due to a combination of many factors, one of the main causes is the difficulty to "make sense" of the phenomena, without learning by first person experiencing and contextualizing in "real-life referents" (Dede, Salzman, Loftin, & Sprague, 1999), hence leading to what is commonly referred to as the abstract nature (Chabay & Sherwood, 2006) of learning physics.

In traditional classrooms, students could be asked to imagine idealized frictionless surfaces where collision carts "move without a decrease in velocity" (Boblick, 1972) and collide in either perfectly elastic collisions without any loss of energy or perfectly inelastic collisions where the velocities of the collision carts become immediately the same. These conditions are almost impossible to achieve using real laboratory equipment (Gluck, 2010), thus we argue that computer simulation could be an appropriate substitute for active learning referents, provided simulations are carefully developed (Weiman & Perkins, 2005), used in appropriate context (N. D. Finkelstein et al., 2005), aided with challenging inquiry activities (Adams, Paulson, & Wieman, 2008) and facilitated by teachers who believe (Hsu, Wu, & Hwang, 2007) in the effectiveness of the tool.

Building on open source codes shared by the Open Source Physics (OSP) community like, Francisco's example of "Collision in one dimension" (Esquembre, 2009), Andrew's (Duffy, 2010) One Dimensional Collision Model for game design, and Fu-Kwun's many other examples on NTNUJAVA Virtual Physics Laboratory, we customize an Easy Java Simulation (EJS) (Esquembre, 2004) computer simulation as a virtual laboratory as in Figure 1 (Wee & Esquembre, 2008), that we hope many teachers will find useful.

## II. PHYSICS MODEL

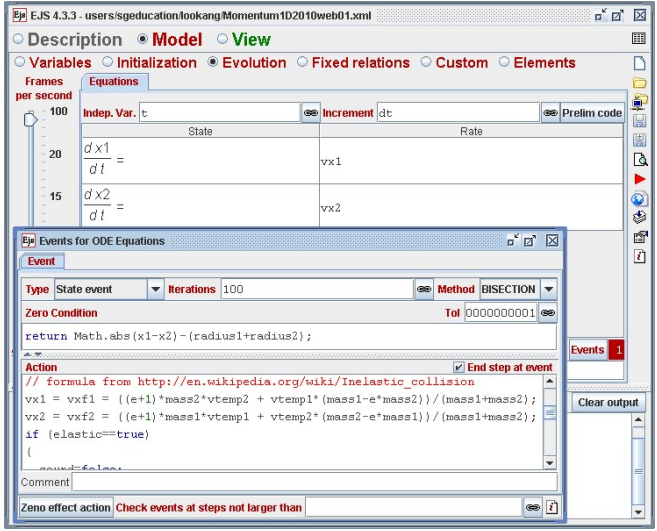

Figure 2. EJS authoring tool view at the 'Evolution Page' showing equations (1) and (2) with the Events for ordinary differential equation (ODE) shown with Type as 'State event', 'Zero Condition' to return to the position where carts are just touching and 'Action' as in equations (3) and (4).

In this simulation, the two-body collision carts model is simulated by both continuous dynamics and discrete transition, where the system dynamics change discretely and the state values jump when the two carts collide. The continuous dynamics is simulated using the 'Evolution Page' as in 0 in EJS by equations (1) and (2), assuming that the $x$ position of the centre of carts 1 and 2 are $x_1$ and $x_2$ respectively and their instantaneous velocities $v_1$ and $v_2$ respectively.

$$\frac{dx_1}{dt} = v_1 \quad (1)$$

$$\frac{dx_2}{dt} = v_2 \quad (2)$$

Notice how easily these equations simulate carts that continue in uniform $x$ direction motion without any loss of energy.

The discrete transition before and after the collision is handled by an event handler in EJS by selecting 'Type = State event' that returns the carts to the position just when the two carts are in contact. The corresponding 'Action' to execute during this event is to assign the physical velocities back to the carts as determined by assuming the concept of coefficient of restitution $e$. The equations for the final velocities $v_1$ and $v_2$ of the carts of masses $m_1$ and $m_2$ as functions of initial velocities $u_1$ and $u_2$ are in equations (3) and (4) respectively.

$$v_1 = \frac{m_1 u_1 + m_2 u_2 - m_2 e(u_1 - u_2)}{(m_1 + m_2)} \quad (3)$$

$$v_2 = \frac{m_1 u_1 + m_2 u_2 + m_1 e(u_1 - u_2)}{(m_1 + m_2)} \quad (4)$$

Due to assumption of coefficient of restitution $e$ given by equation (5) combining with equations (3) and (4), the simulation model allows students to discover that the total momentum of cart 1 and 2 just before and after collision is always the same as shown commonly by equation (6).

$$e = -\frac{v_1 - v_2}{u_1 - u_2} \quad (5)$$

$$m_1 u_1 + m_2 u_2 = m_1 v_1 + m_2 v_2 \quad (6)$$

For the total kinetic energies of cart 1 and 2, kinetic energy loss $E_{Loss} = 0$ for perfectly elastic collision but $E_{Loss} > 0$ in partially or perfectly inelastic collisions as given by equation (7):

$$\frac{1}{2} m_1 u_1^2 + \frac{1}{2} m_2 u_2^2 = \frac{1}{2} m_1 v_1^2 + \frac{1}{2} m_2 v_2^2 + E_{Loss} \quad (7)$$

This Physics model when implemented in a simulation allows experiencing and 'messing about' productively (N. D. Finkelstein, et al., 2005, pp. 010103-010107) serving as a powerful referent tool for learning.

## III. THREE PEDAGOGICAL DESIGN CONSIDERATIONS

To add to the body of knowledge surrounding why simulations could be more effective tools compared to real laboratory equipment (N. D. Finkelstein, et al., 2005; Weiman & Perkins, 2005), we share three pedagogical design insights though not exhaustive, that emerged as being able to push the effectiveness of the tool to be even more useful to students. Teachers interested in other useful features in simulation design could refer to Physics Education Technology (PhET) at University of Colorado (N. Finkelstein, Adams, Keller, Perkins, & Wieman, 2006).

### A. Consistent simulation world view with pen paper representation

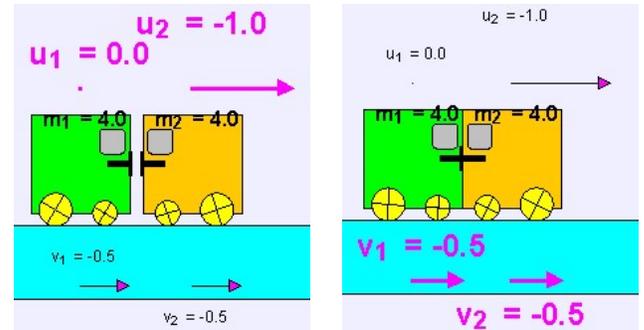

Figure 3. Side world view of the collision carts 1 (green) and 2 (orange) with consistent right pointing velocities representation as those in a typical classroom whiteboard of velocity vectors diagram before (left) and after collision (right).

We realized that the side view of the real equipment setup is easy for students to associate with reality, especially with clear color association (Green for cart 1, Orange for cart 2 for example) and rotating wheels to represent motion.

Customization of simulation with vector representations of velocities $u_1$, $u_2$, $v_1$ and $v_2$ of traditional whiteboard workings with velocity vectors to always point to the right as positive is implemented with appropriate emphasis via an increase in font size and color change, allowing students to relate to classroom discussion of problems as in Figure 3. Several simulations either don't represent the velocities (Christian, 1999) or show pointing to the left (Dubson et al., 2009) while showing a negative magnitude (Hwang, 2010) or cannot simulate negative velocities (Fendt, 2003). Thus, a customized simulation consistent with our pen paper representations by pointing to the right even with negative values minimizes confusion especially when substituting values in equation (6) and (7).

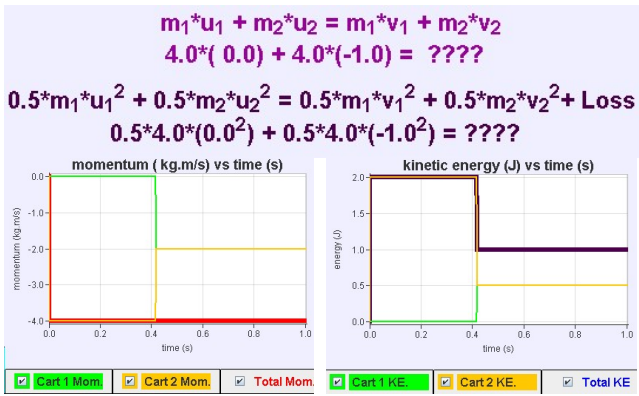

B. *Table of data, scientific graphs and symbolic mathematical representations*

Figure 4. Table of data showing individual and total momentum and kinetic energies before and after collision for ease of retrieving information for data collection and trend analysis.

Figure 5. Scientific graphical plots of momentum (left) notice, the (total momentum) red line is constant at – 4.0 kg.m/s and kinetic energies (right) (total kinetic energies) purple line reflect a drop in kinetic energies at time of collision. There are selectable check-boxes for individual cart and total for multiple representations and further data analyis.

Figure 6. Symbolic mathematical representation of the conservation of momentum (top) and total kinetic energies (bottom) which shows "????" before collision to elicit predictive thinking.

We found that the table of data representation arrange in columns 'Cart 1', 'Cart 2' and 'Total' with rows of before and after of momentum and kinetic energy as in Figure 4 to be useful in our attempts to use the simulation in guided inquiry approach to physics learning. The before and after quantities to be studied in the inquiry laboratory can be easily referred to. This helps to promote productive (N. D. Finkelstein, et al., 2005) activities, as most other simulations (Christian, 1999; Dubson, et al., 2009; Fendt, 2003; Hwang, 2010) need to record at different times the before and after quantities. Graphical representation of momentum and kinetic energy are plotted versus time to give students a time dependent visualization of the physics quantities to be studied as in Figure 5. This graph shows for the case of perfectly inelastic collision ($e = 0$) how total momentum is the same before and after collision (red line) and the total kinetic energies register an energy loss $E_{Loss} > 0$ after collision at time ≈ 0.4 s (purple line). The ability to select cart 1, cart 2 or both allows for clearer visualization and sense making. Symbolic mathematical representation in the form of hints have been added to elicit predictive thinking from students to get them to deduce the curriculum learning outcomes of conservation of momentum and associated kinetic energies as in Figure 6.

In other words, these multiple forms of representations aim to give the students a means to make sense coherently by representing the same phenomena through table of data, scientific graphs and symbolic mathematical representations, supporting the world view in earlier part *A*.

C. *Game for concept testing*

Figure 7. In the game mode, students can test their understanding and receive feedbacks on their answers from the simulation, the level of customized feedback is dependent on the teachers' customized inputs.

To further enhance the engagement and interactivity through the simulation, we suspect that a simple game or puzzle (Adams et al., 2008) could be fun for the students by means of input fields and feedback text. Here students decide which input field to key in first. There could be hints like "no! hint $v_1=v_2$" for perfectly inelastic collision or "no! out by 2.0" and if the answers are correct, the feedback could be "yes! $m_1u_1+m_2u_2 = m_1v_1+m_2v_2$" to reaffirm the students' attempt at the problem game as in Figure 7.

IV. STRENGTHS OF PHYSICAL WORLD LEARNING TO AUGMENT SIMULATION VIRTUAL LEARNING

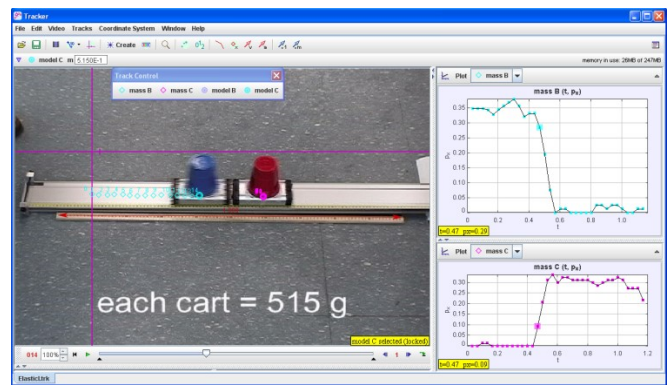

Figure 8. Use Tracker to analyze an elastic collision of a moving cart with a stationary cart of equal mass from Bryan's website that shows the random errors in measurement of momentum $px$ for both masses and the validity of conservation of momentum using velocities just before and just after collision.

Although simulations have its strengths for extending and promoting learning, we would like to present a more balanced view on learning by connecting and suggesting when real physical world setup should be more fore-grounded, augmented by computer simulation.

Tacit 3D experience to learn through physical world collision carts setup, provides very valuable experience to students especially for performance in the real world (Tarekegn, 2009).

Random errors in the measurement of the quantities like velocities; which is part of any scientific measurement, absent in our simulation, is another example of this complementary augmentation on the strengths of physical world learning. For example as in Figure 8, using Tracker (Brown, 2011) to analyze real physical world video such as the ones taken by students themselves and others like Bryan's (Bryan, 2011) video, notice the scientific graphs on the right shows the momentum of the two masses as having statistical fluctuations as random error in measurement of momentum clearly. In addition, the ease of conceptual realization of the significance of the velocities just before and just after, in the verification of conservation of momentum, are valuable experiences when using real equipment, currently not designed in our simulation.

In this balanced view, we suggest possible situations where strengths of physical world learning should be more relied on while other potential weaknesses complimented by strengths afforded by virtual simulations, maintains a connected experience to learning in physical world augmented with simulations.

## V. FEEDBACK FROM STUDENTS

Table 1. Survey results for a 1 hour 40 minutes hands-on lesson with guided inquiry worksheets (N = 64)

|  | 1 Very much (%) | 2 | 3 | 4 | 5 Very little (%) |
|---|---|---|---|---|---|
| 1 How much do you know about the physics taught today before lesson? | **0** | **5** | 23 | 44 | **28** |
| 2 How much do you know about the physics taught today after the lesson? | **6** | **50** | 41 | 0 | **3** |
| 3 I enjoyed learning physics through this activity | **8** | **45** | 34 | 11 | **2** |
| **4 Rate the lesson as a valuable learning experience** | **14** | **58** | **25** | **2** | **2** |

Feedback about the lesson using an earlier version of the simulation is summarized in Table 1. Survey questions 1 and 2 indicate an increase from 5% to 56% of the students perceiving themselves to know very much (scale 1) and much (scale 2) more than before the lesson and a reduction from 72% to 3% self reporting that they still know little (scale 4) or very little (scale 5) after the lesson.

Survey questions 3 and 4 seem to suggest with more than 50% and more than 70% rated the learning experiences as enjoyable and valuable respectively. We speculate factors such as improving teacher's beliefs in effectiveness of technology supported lesson, level of experience in facilitating lessons with use of technology (Wu, Hsu, & Hwang, 2008) and a more challenging guided inquiry task to be able to bring the percentages higher for future lessons.

We include excerpts from the qualitative survey results and informal interviews with the students to give some themes and insights into the conditions and processes during the laboratory lessons. Words in brackets [] are added to improve the readability of the qualitative interviews.

*1) Active learning can be Fun*

"…[It] is an eye opener...[we] don't usually get to learn with virtual learning environment…and it makes learning fun and interesting".

"The lesson was fun and makes us think instead of just listen[ing] to teacher and remember[ing] whatever the teacher said".

"It makes learning much more interesting and fun. It makes us want to learn and find out more about the topic".

*2) Need experience to understand*

"…it [this lab] lets me figure out the concepts rather than just listen[ing] and believing what is taught without understanding".

"Normally people would have to experience any physics concepts themselves through hands[-]on to really remember concepts. Lectures on the other hand may not be effective since maybe what the lecturer is bringing through us is unclear, and thus practical lessons to learn concepts is a great learning deal".

*3) Simulation can support inquiry learning and thinking like real scientist*

"These kinds of lesson force us to think critically. It makes us look at the results, analyze and then find the trend within, which is a really good way to learn independently. It also gives us confidence and a sense of accomplishment when the conclusions we arrive at are correct."

"Such vlab[virtual lab] lesson effectively utilizes the IT[information technology] resources to enhance lessons, making physics lessons less dry. Besides, by identifying trends in values first hand, I can remember it easier rather than via lecture notes and slides"

*4) Need for strong inquiry learning activities*

"The activity worksheet did not generate much thinking and concept understanding, just simply presents a set of values to copy to get the answers".

"It [virtual lab] helps hasten the process of learning but the exchange of data [in the worksheet activities] is troublesome".

*5) Need for testing and well designed simulation (N. D. Finkelstein, et al., 2005)*

Some students suggest visual and audio enhancements like "better quality so that the simulations could be more interesting and appealing" and "add sound effects".

A good suggestion surface is to make the "program[simulation] designed as a game , thereby making it more interactive. At the end a table can be provided and it would provide us[students] with the values. From there, we do analysis".

This suggestion has inspired us to design 'C Game for concept testing' in earlier part III.

*6) Appreciative learners*

"I[student] really thank you for spending time coming up with this program[simulation]. You are really an educator who cares and dares to try new things. Thanks! Hope you can come up with even better programs so that they can empower students in physics subject."

"Thank you teachers for spending time to develop this app[lication] :)"

## VI. Conclusion

The theoretical physics model of a two-body collision system in one dimension is discussed and implemented in EJS and the equations (1) to (7) should be applicable to any modeling tool such as VPython (Scherer, Dubois, & Sherwood, 2000) or Modellus (Teodoro, 2004). Three pedagogical considerations in simulation design such as 1) consistent simulation world view with pen paper representations, 2) data table, scientific graphs and symbolic mathematical representations for ease of data collection and multiple representational visualizations (Ainsworth, 2008) and 3) game for simple concept testing are implemented in our simulation that we believe can further support learning.

We also suggest using physical world setup to be augmented complimentarily with simulation while highlighting three advantages of real collision carts equipment like tacit 3D experience, random errors in measurement and conceptual significance of conservation of momentum applied to just before and after collision.

General feedback from the students has been relatively positive, triangulated from the survey questions, interviews with students and discussions with teachers and we hope teachers will find the simulation useful in their own classes.

## Acknowledgement

We wish to acknowledge the passionate contributions of Francisco Esquembre, Fu-Kwun Hwang and Wolfgang Christian for their ideas and insights in the co-creation of interactive simulation and curriculum materials.

## Reference

Adams, W. K., Paulson, A., & Wieman, C. E. (2008, July 23-24). *What Levels of Guidance Promote Engaged Exploration with Interactive Simulations?* Paper presented at the Physics Education Research Conference, Edmonton, Canada.

Adams, W. K., Reid, S., Lemaster, R., McKagan, S. B., Perkins, K. K., Dubson, M., & Wieman, C. E. (2008). A Study of Educational Simulations Part 1 -- Engagement and Learning. [Article]. *Journal of Interactive Learning Research, 19*(3), 397-419.

Ainsworth, S. (2008). The Educational Value of Multiple-representations when Learning Complex Scientific Concepts. In J. K. Gilbert, M. Reiner & M. Nakhleh (Eds.), *Visualization: Theory and Practice in Science Education* (Vol. 3, pp. 191-208). University of Nottingham, UK: Springer Netherlands.

Boblick, J. M. (1972). Discovering the Conservation of Momentum Through the Use of a Computer Simulation of a One-Dimensional Elastic Collision. [Article]. *Science Education, 56*(3), 337-344.

Brown, D. (2011). Tracker Free Video Analysis and Modeling Tool for Physics Education Retrieved 20 October, 2010, from http://www.cabrillo.edu/~dbrown/tracker/

Bryan, J. A. (2011). Video Analysis: Real World Investigations for Physics and Mathematics. *Elastic Collisions I moving cart collides with stationary cart of equal mass*, from http://jabryan.iweb.bsu.edu/videoanalysis/index.htm

Chabay, R., & Sherwood, B. (2006). Restructuring the introductory electricity and magnetism course. *American Journal of Physics, 74*(4), 329-336.

Christian, W. (1999). Elasticity of a Collision [application/Java 1.1 Physlets]. USA: Open Source Physics. Retrieved from http://webphysics.davidson.edu/physlet_resources/bu_semester1/c13_elasticity.html

Dede, C., Salzman, M., Loftin, R., & Sprague, D. (1999). Multisensory Immersion as a Modeling Environment for Learning Complex Scientific Concepts. In W. Feurzeig & N. Roberts (Eds.), *Modeling and simulation in science and mathematics education* (Vol. 1). New York: Springer.

Dubson, M., Loeblein, T., Perkns, K., Gratny, M., Olson, J., & Reid, S. (2009). PhET: Collision Lab (Version Version: 0.01 (45337) Build Date: Oct 22, 2010): University of Colorado. Retrieved from https://phet.colorado.edu/en/simulation/collision-lab

Duffy, A. (2010). One Dimensional Collision Model [application/java]. Boston University, USA: Open Source Physics. Retrieved from http://www.compadre.org/Repository/document/ServeFile.cfm?ID=10000&DocID=1637

Esquembre, F. (2004). Easy Java Simulations: A software tool to create scientific simulations in Java. *Computer Physics Communications, 156*(2), 199-204.

Esquembre, F. (2009). Collision in one dimension [application/java]. Universidad de Murcia, Spain: Open Source Physics. Retrieved from http://www.um.es/fem/EjsWiki/Main/ExamplesCollision1D

Fendt, W. (2003). Elastic and Inelastic Collision [application/java]. Stadtbergen, Germany: Fendt, Walter. Retrieved from http://www.walter-fendt.de/ph14e/collision.htm

Finkelstein, N., Adams, W., Keller, C., Perkins, K., & Wieman, C. (2006). High-tech tools for teaching physics: The physics education technology project. *MERLOT Journal of Online Learning and Teaching, 2*(3), 110 - 120.

Finkelstein, N. D., Adams, W. K., Keller, C. J., Kohl, P. B., Perkins, K. K., Podolefsky, N. S., . . . LeMaster, R. (2005). When Learning about the Real World is Better Done Virtually: A Study of Substituting Computer Simulations for Laboratory Equipment.

*Physical Review Special Topics - Physics Education Research, 1*(1), 010103.

Gluck, P. (2010). Elastic and Inelastic Collisions. *The Physics Teacher, 48*(3), 158-158. doi: 10.1119/1.3317446

Hsu, Y.-S., Wu, H.-K., & Hwang, F.-K. (2007). Factors Influencing Junior High School Teachers' Computer-Based Instructional Practices Regarding Their Instructional Evolution Stages. *Educational Technology & Society, 10*(4), 118-130.

Hwang, F.-K. (2010). 1D collision : Conservation of Momentum [application/java JDK]. Taiwan: National Taiwan Normal University. Retrieved from http://www.phy.ntnu.edu.tw/ntnujava/index.php?topic=5.0

Scherer, D., Dubois, P., & Sherwood, B. (2000). VPython: 3D interactive scientific graphics for students. *Computing in Science and Engg., 2*(5), 56-62. doi: 10.1109/5992.877397

Singh, C., & Rosengrant, D. (2003). Multiple-choice test of energy and momentum concepts. *American Journal of Physics, 71*(6), 607-617. doi: 10.1119/1.1571832

Tarekegn, G. (2009). Can computer simulations substitute real laboratory apparatus? *Lat. Am. J. Phys. Educ. Vol, 3*(3), 506.

Teodoro, V. D. (2004). Playing Newtonian Games with Modellus. *Physics Education, 39*(5), 421-428.

Wee, L. K., & Esquembre, F. (2008). Ejs open source java applet 1D collision carts Elastic and Inelastic Collision [application/java]. Singapore: Open Source Physics. Retrieved from http://www.phy.ntnu.edu.tw/ntnujava/index.php?topic=831.0

Weiman, C., & Perkins, K. (2005). Transforming Physics Education. *Physics Today, 58*(11), 36-40.

Wu, H.-K., Hsu, Y.-S., & Hwang, F.-K. (2008). Factors Affecting Teachers' Adoption of Technology in Classrooms: Does School Size Matter? *International Journal of Science and Mathematics Education, 6*(1), 63-85. doi: 10.1007/s10763-006-9061-8

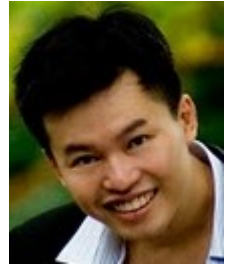

AUTHOR

Loo Kang WEE is currently an educational technology officer at the Ministry of Education, Singapore and a PhD candidate at the National Institute of Education, Singapore. He was a junior college physics lecturer and his research interest is in designing simulations for physics learning. His curriculum materials and simulation models can be accessed from the NTNUJAVA Virtual Physics Laboratory and Open Source Physics Collection under Creative Commons Attribution License.